\documentclass[twocolumn,aps]{revtex4}

\usepackage{amsmath}

\usepackage{amssymb}
\usepackage{graphicx}

\begin{document}

\title{Robust transmission of non-Gaussian entanglement over optical fibers}
\author{Asoka Biswas$^1$ and Daniel A. Lidar$^{1,2}$}
\affiliation{$^1$Department of Chemistry, University of Southern California, Los Angeles,
CA 90089\\
$^2$Departments of Electrical Engineering-Systems and Physics, University of
Southern California, Los Angeles, CA 90089}
\date{\today}

\begin{abstract}
We show how the entanglement in a wide range of continuous variable
non-Gaussian states can be preserved against decoherence for long-range
quantum communication through an optical fiber. We apply protection via
decoherence-free subspaces and quantum dynamical decoupling to this end. The
latter is implemented by inserting phase shifters at regular intervals $%
\Delta $ inside the fiber, where $\Delta $ is roughly the ratio of the speed of light in the fiber
to the bath high-frequency cutoff. Detailed estimates of relevant parameters are provided using the
boson-boson model of system-bath interaction for silica fibers, and $\Delta $ is found to be on the
order of a millimeter.
\end{abstract}

 \maketitle

\section{Introduction}

Gaussian entangled states of two subsystems are well studied in the quantum
communication and information literature. These states are often encountered
in quantum communication experiments. However non-Gaussian entangled states
are also important in the context of quantum communication. For example,
such a communication system may be constructed in two steps \cite{geo}: (a)
encoding based on product states of Gaussian states, (b) decoding based on
measurement on this continuous set of Gaussian states. During decoding, one
must generate superpositions of input Gaussian states, which are essentially
non-Gaussian states. Thus quantum communication systems may require
non-Gaussian states. Moreover, it is impossible to achieve a quantum speed-up
using only harmonic oscillators and corresponding Gaussian operations \cite%
{bartlett}: the dynamics of such a system can be efficiently simulated
classically. This, in turn, means that to gain a quantum advantage in this
scenario, one needs to use non-Gaussian operations. It has also been shown
that the distillation of entanglement from two Gaussian entangled states is
impossible using only local Gaussian operations and classical communication
\cite{eisert}. Thus, both quantum speed-up and distillation of entanglement,
which have a close relationship with quantum error correction, require
non-Gaussian operations. Recently there have been a few proposals to detect
entanglement in such non-Gaussian states \cite{biswas,vogel,hillery}.

In this work we focus on the problem of preserving non-Gaussian entanglement in
noisy quantum communication channels. There have been several proposals
based on quantum purification protocols and quantum repeaters to communicate
entangled photonic qubits over long distances \cite{bennett,deutsch,dur}.
Alternatively, entanglement between distant nodes can be prepared by
measurements along a chain of intermediate particles \cite%
{WuLidarSchneider:04,Compagno:04}. The problem we address here is quite
different: rather than setting up remote entanglement between distant nodes,
we consider the problem of transmitting entangled field states for a long
distance along an optical fiber. Specifically, we propose a method that
preserves the multiphoton entanglement of a class of non-Gaussian states
transmitted through an optical fiber. Most optical fibers are known to have
minimum loss windows at wavelengths of the order of few microns. The optical
frequencies generally suffer attenuation inside such a fiber. This
loss adds to other decoherence processes which degrade the fidelity of
entanglement transmission. We propose a
hybrid approach to control decoherence of optical-frequency non-Gaussian
states inside an optical fiber. Specifically, we utilize decoherence-free
subspace (DFS) \cite{dfs} and quantum dynamical bang-bang (BB) decoupling
\cite{BB} protection to preserve non-Gaussian entanglement. The DFS
protection is used against differential dephasing of the two field modes
used to construct non-Gaussian entangled states. The BB process is applied
in order to deal with the remaining relevant decoherence sources, in
particular Raman scattering. This is done by inserting phase shifters at
regular intervals along the length of the optical fiber, similarly to the
proposal in \cite{lidar} where this spatial BB procedure was used to protect
single photon polarization states transmitted through optical fibers. In
this manner we provide a novel application of the general hybrid DFS-BB
strategy proposed in \cite{byrd_lidar} (see also \cite{LidarWu-ERD}).

The structure of the paper is as follows. In Sec.~\ref{int}, we introduce a class of non-Gaussian
states and model their interaction with an optical fiber. In Sec.~\ref{hybrid}, we describe in
detail a hybrid approach to eliminating their decoherence during transmission through an optical
fiber. In Sec.~\ref{inhom}, we provide a numerical estimate of loss of entanglement through a
realistic fiber using a boson-boson model of interaction between the field and the fiber.

\section{Interaction mechanism of non-Gaussian entangled states with fiber}

\label{int}

\subsection{A class of two-mode non-Gaussian states and their entanglement}

The simplest examples of non-Gaussian states of the electromagnetic field are the single
photon states. Other examples are states generated by excitations of a
Gaussian state \cite{tara_thermal,gsapuri}. Another method to produce
non-Gaussian states uses state-reduction \cite{gsa_qo,yamamoto,bellini}. A
recent experimental proposal discusses how to generate non-Gaussian states
by subtracting a photon from each mode of a two-mode squeezed vacuum state
\cite{grangier}. We consider non-Gaussian entangled states $|\psi\rangle$ of
two field modes $a$ and $b$ ($a$ and $b$ are bosonic annihilation
operators), produced by subtracting $p (>0)$ photons from one of the modes
(say, $a$) of a squeezed vacuum state $|\phi\rangle$, given by,
\begin{eqnarray}  \label{state}
|\psi\rangle&=&a^p|\phi\rangle\equiv \frac{1}{P}\sum_{n=0}^{\infty}\sqrt{%
\frac{(n+p)!}{n!}}\zeta^{n+p}|n,n+p\rangle\;, \\
|\phi\rangle &\equiv&\sum_{n=0}^{\infty}\zeta^n|n,n\rangle,
\end{eqnarray}
where $\zeta$ is the complex squeezing parameter and $P$ is the
normalization constant. This state is non-Gaussian in the sense that its
coordinate representation is a non-Gaussian function of space-coordinates $x$
and $y$, as given by,
\begin{equation}
\psi(x,y)=\frac{\zeta^p}{\sqrt{2^p\pi}}\sum_{n=0}^\infty\frac{(\zeta/2)^n}{n!%
}H_n(x)H_{n+p}(y)e^{-(x^2+y^2)/2},
\end{equation}
where $H_n(x)$ is the $n$th order Hermite polynomial. Note that for $p=0$, the
state becomes Gaussian.

The entanglement in this state can be verified by the Peres-Horodecki
criterion \cite{peres}. We find that the density matrix of the state $%
|\psi\rangle$ under transpose of the $b$ mode transforms into
\begin{eqnarray}
\sigma&=&\frac{1}{P^2}\sum_{n,m=0}^\infty \zeta^{n+p}\zeta^{*^{m+p}}\sqrt{%
\frac{(n+p)!(m+p)!}{n!m!}}  \nonumber \\
&&|n,m+p\rangle\langle m,n+p|\;.
\end{eqnarray}
The eigenvalues of the above matrix can be calculated as
\begin{eqnarray}
\lambda_{nn}&=&\frac{1}{P^2}|\zeta|^{2(n+p)}\frac{(n+p)!}{n!}\;\;\;\forall
n\;,  \nonumber \\
\\
\lambda_{nm}&=&\pm\frac{1}{P^2}|\zeta|^{n+m+2p}\sqrt{\frac{(n+p)!(m+p)!}{n!m!%
}}\;\;\;\forall n\neq m\;.  \nonumber
\end{eqnarray}
Existence of negative eigenvalues of the matrix $\sigma$ reflects that the
state $|\psi\rangle$ is entangled. The negativity $\mathcal{N}$ \cite{vidal}
of the state $|\psi\rangle$ can be written as
\begin{equation}  \label{neg}
\mathcal{N}=\frac{1}{P^2}\sum_{n=0}^\infty\sum_{m\neq
n}^\infty|\zeta|^{n+m+2p}\sqrt{\frac{(n+p)!(m+p)!}{n!m!}}\;,
\end{equation}
which is the absolute sum of all negative eigenvalues of the density matrix $%
\sigma$. The non-zero value of $\mathcal{N}$ reflects that the state is
entangled. Deviation of the value of $\mathcal{N}$ from zero is a measure of
the degree of entanglement. It is thus clear from Fig.~1 that the two modes
become more entangled with increasing values of $|\zeta|$.
\begin{figure}[tbp]
\scalebox{0.25}{\includegraphics{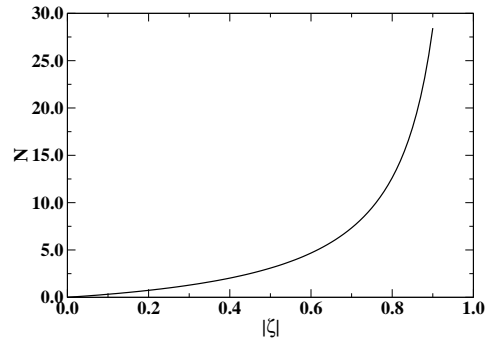}}
\caption{Variation of negativity $\mathcal{N}$ with the squeezing parameter $%
|\protect\zeta|$ for $p=1$.}
\label{fig1}
\end{figure}

\subsection{Model of interaction with an optical fiber}

We assume that the state $|\psi \rangle $ is transmitted through an optical
fiber of length $L$ and thereby interacts with the fiber modes, which leads
to decoherence of the state. An optical fiber consists of many dielectric
molecules, each of which contains many electronic energy levels. The
interaction of the input field states with the fiber can be reasonably
described by a molecule-field interaction Hamiltonian. Dominant
decoherence mechanisms are phase damping and energy exchange between
the field and the molecules.

The exchange processes can be expressed via the
following Hamiltonian, under the rotating-wave approximation:
\begin{equation}
H_{I}\equiv \sum_{i}(A_{i}^{\dag }B_{i}+\mathrm{h.c.}),
\end{equation}%
where $A_{i}$ describes the field operators and $B_{i}$ describes the
molecular operators. Here the molecules in the fiber act as the bath,
leading to decoherence.

The exact form of the field operators $A_{i}$ in the above Hamiltonian
depends upon the model of the interaction. For example, if the field modes
are at near-resonance with the single-photon transition in the molecules,
dipole coupling between them leads to $A_{i}\equiv a,b$. Dipole
coupling dominates higher order coupling (e.g.,
quadrupole coupling or magnetic dipole coupling), which would lead to
multi-photon processes. Thus, at near-resonance, single-photon absorption by
the molecules leads to decoherence. Clearly, far from single-photon
resonance, higher order processes can dominate. In light of these
considerations, we assume that the frequencies of the two field modes are so
chosen that two-photon processes may occur in the system, while the
cross-section of single-photon processes becomes negligible. Specifically,
we choose the frequencies of the field-modes to be much smaller than the
energy gaps between the ground and first excited electronic states.

There
are two different kinds of two-photon processes that may occur in a
molecule: annihilation (creation) of two photons described by $a^{2}$, $%
b^{2} $, and $ab$ ($a^{\dag 2}$, $b^{\dag 2}$, and $a^{\dag }b^{\dag }$),
and photon-number conserving processes (described by $a^{\dag }b$, $ab^{\dag
}$, $a^{\dag }a$, and $b^{\dag }b$). If we assume the molecules are in their
electronic ground states (at low temperature, as discussed later), the
process of absorption of two photons of two orthogonal modes is disallowed
due to certain selection rules \cite{cohen}. Moreover, as both modes
propagate in the same direction through the fiber, Doppler shift of the
photon frequency causes the absorption processes of two
photons in the same mode to be off-resonance \cite{cohen}. Thus, we are led to a situation
where most of the molecular levels are resonant with the second kind of
two-photon transitions. This means that low-energy scattering of the
photons, namely Raman (described by $a^{\dag }b$ and $ab^{\dag }$) and
dephasing processes (described by the number operators $n_{a}\equiv a^{\dag
}a$ and $n_{b}\equiv b^{\dag }b$, which is equivalent to Rayleigh
scattering), are most likely to occur inside the fiber. In this case, we can
write the effective interaction Hamiltonian in the rotating wave
approximation as
\begin{eqnarray}
H_{I} &=&\sum_{i}g_{i}(a^{\dag }bB_{i}+ab^{\dag }B_{i}^{\dag
})+\sum_{i}(\Gamma _{a}^{i}n_{a}+\Gamma _{b}^{i}n_{b})B_{\mathrm{ph}}^{i}\;
\nonumber \\
&\equiv &H_{I}^{(1)}+H_{I}^{(2)},  \label{ham}
\end{eqnarray}%
where $B_{i}$ is the annihilation operator that corresponds to relevant
transitions of $i$th molecule in the fiber, $g_{i}$'s are the coupling
strengths, $B_{\mathrm{ph}}^{i}$ is the dephasing operator for the $i$th
molecule (the exact form of this will be discussed later), and $\Gamma
_{a}^{i}$ and $\Gamma _{b}^{i}$ are the respective dephasing rates of the
two modes $a$ and $b$.

\section{A hybrid approach to eliminate decoherence}

\label{hybrid}

\subsection{A decoherence-free subspace against differential dephasing}

In this section we discuss a hybrid approach to suppress these two-photon
processes, i.e., how the effect of the interaction Hamiltonian (\ref{ham})
on the state $|\psi \rangle $ can be eliminated. We start with the dephasing
processes described by the number operators $n_{a}\equiv a^{\dag }a$ and $%
n_{b}\equiv b^{\dag }b$. We rewrite the dephasing part of the Hamiltonian (%
\ref{ham}), $\sum_{i}(\Gamma _{a}^{i}n_{a}+\Gamma _{b}^{i}n_{b})B_{\mathrm{ph%
}}^{i}$, in terms of two collective operators
\begin{equation}
A_{\pm }=n_{a}\pm n_{b}
\end{equation}%
as $\sum_{i}(\Gamma _{-}^{i}A_{-}+\Gamma _{+}^{i}A_{+})B_{\mathrm{ph}}^{i}$,
where $\Gamma _{\pm }^{i}=(\Gamma _{a}^{i}\pm \Gamma _{b}^{i})/2$. Note that
the two-mode number states $|n,n+p\rangle $ and $|n+p,n\rangle $ are
eigenstates of the operator $A_{-}=n_{a}-n_{b}$ for all integers $p\geq 0$.
Thus these states form manifolds of decoherence-free subspaces with respect
to differential dephasing for a given $p$. In any manifold $\{|n,n+p\rangle
\}$ ($p$ fixed), any arbitrary superposition of all possible states is also
a decoherence-free state \cite{dfs}. In this way, in the present case the
state $|\psi \rangle $ is a DFS under the action of collective dephasing $%
A_{-}$. Note that this protection does not require encoding, in contrast to
the usual construction of decoherence-free subspaces \cite{dfs}.

\subsection{Bang-bang decoupling of the Raman process}

As a second layer of protection of the state against other decoherence
processes (described by Raman interactions), we now follow the general
hybrid DFS-BB method proposed in \cite{byrd_lidar}. In the standard BB
decoupling methods \cite{BB}, one uses very short pulses so as to cancel
the effective interaction Hamiltonian. Thus one ends up with only that
component of the total Hamiltonian of system and bath which commutes
with the BB pulses. If the pulses are appropriately chosen,
entanglement generation between the system and bath states can thus be
prevented, and decoherence of the system is prevented.
However, one has to apply the pulses in intervals shorter than the
timescale of decay of the bath correlation. A recent proposal
\cite{lidar} uses a spatial,
rather than temporal version of this idea to overcome decoherence of a
single-photon polarization state in optical fibers. Ref.~\cite{lidar}
shows how to replace the short time-dependent pulses with phase shifters at
regular intervals. Our approach here is similar except that we deal with
non-Gaussian entanglement rather than single-photon states, and introduce a
hybrid DFS-BB approach. In this way \emph{entanglement can be transmitted
  over a long distance in quantum communication systems}.

Let us start with
the total system-bath Hamiltonian in the form
\begin{eqnarray}
H &=&H_{0}+H_{I},\;  \nonumber \\
H_{0} &=&\hbar \omega _{a}n_{a}+\hbar \omega _{b}n_{b}+H_{B},
\end{eqnarray}%
where $\omega _{a,b}$ are the frequencies of the two field modes, $H_{B}$ is
the free Hamiltonian of the molecular bath, and $H_{I}$ is given by Eq. (\ref%
{ham}). The frequencies $\omega _{a,b}$ are chosen so that they are
off-resonant with single-photon transitions, but possibly resonant with
Raman transitions in the fiber molecules. Here and below we omit the
zero-point energies when writing oscillator Hamiltonians.

The initial state is $\left\vert \Psi _{0}\right\rangle =|\psi \rangle
\left\vert M\right\rangle $, where $|\psi \rangle $ is the input
non-Gaussian entangled state and $\left\vert M\right\rangle $ is the state
of the bath (for simplicity we use a pure-state notation also for the bath;
below we consider the effect of the bath's state more carefully). At the
time $\tau _{L}=L/v$ -- where $L$ is the length of the fiber, and $v$ is the
average speed of light in the fiber -- the total wave function is $%
\left\vert \Psi (\tau _{L})\right\rangle =U_{\mathrm{f}}(\tau _{L},0)$ $%
\left\vert \Psi _{0}\right\rangle $ (the f subscript stands for
\textquotedblleft free evolution\textquotedblright ). Here the exact
normal-ordered propagator is $U_{\mathrm{f}}(\tau _{L},0)=\,:\exp
[-i\int_{0}^{L}[H_{I}(x)+H_{0}(x)]dx]:$ (in units where $\hbar =1$). Now,
let $\tau =\Delta /v$, and let $N=L/\Delta \gg 1$ so that we can expand the
propagator as $U_{\mathrm{f}}(\tau _{L},0)\approx e^{-i\overline{H}(N\Delta
)\tau }\cdots e^{-i\overline{H}(2\Delta )\tau }e^{-i\overline{H}(\Delta
)\tau }$, where%
\begin{eqnarray}
\overline{H}(k\Delta ) &=&\frac{1}{\Delta }\int_{(k-1)\Delta }^{k\Delta
}[H_{0}(x)+H_{I}(x)]dx  \nonumber \\
&\equiv &\overline{H_{0}}(k\Delta )+\overline{H_{I}}(k\Delta )
\end{eqnarray}%
is the average Hamiltonian over the $k$th segment. I.e., we have neglected
deviations from average fiber homogeneity, $\delta _{k}=\langle (H(k\Delta
)-[H_{I}(k\Delta )+H_{0}(k\Delta )])^{2}\rangle $ (we discuss such
deviations in Section \ref{inhom}). The spatial dependence of $H_{0}$ can
come from the bath self-Hamiltonian $H_{B}$, while that of $H_{I}$ can come
from the coefficients $g_{i}$ and $\Gamma _{a,b}^{i}$ [Eq.~(\ref{ham})].
Therefore:%
\begin{eqnarray}
\overline{H_{0}}(k\Delta ) &=&\hbar \omega _{a}n_{a}+\hbar \omega _{b}n_{b}+%
\overline{H_{B}}(k\Delta )  \nonumber \\
\overline{H_{I}}(k\Delta ) &=&\sum_{i}\overline{g_{i}}(a^{\dag
}bB_{i}+ab^{\dag }B_{i}^{\dag })
\nonumber \\
&&
+\sum_{i}(\overline{\Gamma _{a}}n_{a}+\overline{\Gamma _{b}}n_{b})B_{\mathrm{ph}}^{i}  \nonumber \\
&\equiv &\overline{H_{I}^{(1)}}(k\Delta )+\overline{H_{I}^{(2)}}(k\Delta ).
\end{eqnarray}%
Moreover, in subsection \ref{colldep} below we argue that $H_{B}$ is
effectively a molecular oscillator Hamiltonian, so that $\overline{H_{B}}%
(k\Delta )=\sum_{s}\hbar \overline{\Omega _{s}}s$, where $\overline{\Omega
_{s}}=\frac{1}{\Delta }\int_{(k-1)\Delta }^{k\Delta }\Omega _{s}(x)dx$ is an
average frequency in an $s$-phonon state.

To eliminate the Raman processes dynamically, we propose use of phase
shifters, defined by the following operator
\begin{equation}
\Pi =e^{i\pi (n_{a}-n_{b})/2}=\Pi ^{\dag },
\end{equation}%
which generates a relative phase of $\pi $ between the two modes
(alternatively we can define $\Pi $ as $e^{i\pi n_{a}}$ or as $e^{i\pi
n_{b}} $). When these phase shifters are incorporated inside the fiber at
intervals $\Delta $, the Raman interaction part in the Hamiltonian (\ref{ham}%
) effectively vanishes. This occurs because of the following identities.
First, it is simple to show, using the Baker-Campbell-Hausdorff (BCH)
formula \cite{Reinsch:00} that
\begin{equation}
e^{i\phi n_{a}}a^{\dagger }e^{-i\phi n_{a}}=e^{i\phi }a^{\dagger },\quad
e^{i\phi n_{a}}ae^{-i\phi n_{a}}=e^{-i\phi }a,  \label{eq:n-b}
\end{equation}%
and similarly for $n_{b}$ and $b$. Therefore%
\begin{eqnarray}
\Pi a^{\dag }b\Pi ^{\dag } &=&(e^{i\pi n_{a}/2}a^{\dagger }e^{-i\pi
  n_{a}/2})(e^{-i\pi n_{b}/2}be^{i\pi n_{b}/2}) \nonumber \\
&=& -a^{\dag }b\;,
\\
\Pi ab^{\dag }\Pi ^{\dag } &=&(e^{i\pi n_{a}/2}ae^{-i\pi n_{a}/2})(e^{-i\pi
n_{b}/2}b^{\dag }e^{i\pi n_{b}/2}) \nonumber \\
&=& -ab^{\dag }\;. \end{eqnarray}%
\emph{These identities imply that the Raman interaction is effectively
time-reversed every }$2\Delta $\emph{\ due to the action of the phase
shifters}. To show the utility of this result note first that
\begin{equation}
\Pi \overline{H}\Pi =\overline{H_{0}}-\overline{H_{I}^{(1)}}+\overline{%
H_{I}^{(2)}},  \label{eq:flip}
\end{equation}%
because the field term of $H_{0}$ and $H_{I}^{(2)}$ obviously commute with $%
\Pi $ and because averaging commutes with the phase shifter operation. Now,
if we install thin phase-shifters inside the fiber at positions $x=0,\Delta
,2\Delta ,...$, from $A$ to $B$, the evolution will be modified to
\begin{equation}
U(\tau _{L},0)\approx e^{-i\overline{H}(N\Delta )\tau }\cdots \Pi e^{-i%
\overline{H}(2\Delta )\tau }\Pi e^{-i\overline{H}(\Delta )\tau }\Pi .
\label{pulse}
\end{equation}%
Note that in writing this expression we have neglected the variation of $H$
inside the phase-shifter; this will hold provided that the phase-shifter
width is much smaller than the distance over which deviations $\delta _{k}$
from average fiber homogeneity become significant. Now assume that the \emph{%
average} Hamiltonians over two successive segments are equal:
\begin{eqnarray}
\overline{H_{I}^{(i)}}((k+1)\Delta ) &=&\overline{H_{I}^{(i)}}(k\Delta
)\quad i=1,2  \nonumber \\
\overline{H_{0}}((k+1)\Delta ) &=&\overline{H_{0}}(k\Delta ).
\label{eq:approxH}
\end{eqnarray}%
The better this approximation, the better our hybrid DFS-BB\ method will
perform; we address deviations in Section \ref{inhom}. In this case, to
first order in $\tau $, and using Eqs.~(\ref{eq:flip}) and (\ref{pulse}), we
have exact cancellation of $H_{I}^{l}$ between successive segments:
\begin{eqnarray}
&&e^{-i\overline{H}((k+1)\Delta )\tau }\Pi e^{-i\overline{H}(k\Delta )\tau
}\Pi  \nonumber \\
&=&e^{-i\overline{H}((k+1)\Delta )\tau }e^{-i\Pi \overline{H}(k\Delta )\Pi
\tau }  \nonumber \\
&=&e^{-i\tau (\overline{H_{0}}(k\Delta )+\overline{H_{I}^{(1)}}(k\Delta )+%
\overline{H_{I}^{(2)}}(k\Delta ))}\nonumber\\
&&\times e^{-i\tau (\overline{H_{0}}(k\Delta )-%
\overline{H_{I}^{(1)}}(k\Delta )+\overline{H_{I}^{(2)}}(k\Delta ))}
\nonumber \\
&=&e^{-2i(\overline{H_{0}}(k\Delta )+\overline{H_{I}^{(2)}}(k\Delta ))\tau },
\end{eqnarray}%
where the smallness of $\tau $ is the justification for adding the arguments
of the two exponentials in obtaining the last line. Using Eq.~(\ref%
{eq:approxH}) again, this yields the overall evolution operator
\begin{equation}
U(\tau _{L},0)\approx e^{-i\tau _{L}(\overline{H_{0}}(L)+\overline{%
H_{I}^{(2)}}(L))}.
\end{equation}

Thus, to first order in $\tau $ (or the inter-phase shifter distance $\Delta
$), we have eliminated the Raman term $\overline{H_{I}^{(1)}}$, and are left
with the dephasing term $\overline{H_{I}^{(2)}}=\sum_{i}(\overline{\Gamma
_{-}}A_{-}+\overline{\Gamma _{+}}A_{+})B_{\mathrm{ph}}^{i}$, as $\overline{%
H_{0}}$ causes no decoherence. Considering $\overline{H_{I}^{(2)}}$, we note
that the $A_{-}$ component yields only an overall phase on the state $|\psi
\rangle $. In fact, the DFS corresponding to $A_{-}$ remains invariant under
the action of $\Pi $ as $\Pi $ obviously commutes with $A_{-}$. Therefore,
the overall evolution operator reduces to
\begin{equation}
U(\tau _{L},0)\approx e^{-i\tau _{L}(\overline{H_{0}}+\sum_{i}\overline{%
\Gamma _{+}}A_{+}B_{\mathrm{ph}}^{i})},  \label{eq:U}
\end{equation}%
(where averages can be considered as being evaluated at $L$) and our
remaining task is to eliminate the collective dephasing term (proportional
to $A_{+}$). However, it turns out that it is impossible to do this with BB
while using only linear optical elements. Fundamentally, the reason for this
is that the group generated by $\{n_{a}-n_{b},a^{\dagger }b,b^{\dagger }a\}$
(phase shifters and beam splitters) is non-compact, which means that it can
at most apply a dilation but not a sign change, as required for time
reversal in the BB protocol. However, as we argue next, the collective
dephasing term nevertheless has no significant effect on the preservation of
non-Gaussian entanglement.

\subsection{Suppression of collective dephasing}

\label{colldep}

In the low temperature limit, most of the molecules in each segment of
length $\Delta $ reside in their ground electronic states. In thermal
equilibrium, the molecular vibrational states obey Maxwell-Boltzmann
statistics. The state of the molecular bath can be well described by a
density matrix $\rho _{B}=\sum_{s}p_{s}|\phi _{s}\rangle \langle \phi _{s}|$%
, where $p_{s}=\exp (-E_{s}/k_{B}T)/Z$ is the Boltzmann probability of an $s$%
-phonon excitation with energy $E_{s}=\hbar \Omega _{s}$, $k_{B}$ is the
Boltzmann constant, and $Z=\sum_{s}\exp (-E_{s}/kT)$ is the partition
function, and $|\phi _{s}\rangle $ is the $s$-phonon excited state. This is
equivalent to the assumption that the molecular Hamiltonian $H_{B}$ is an
oscillator Hamiltonian, i.e., $H_{B}=\sum_{s}\hbar \Omega _{s}s$ ($s$ is the
phonon number operator), as mentioned in the previous subsection. Moreover,
we assume that the field couples to the molecular vibrations. Then $%
[H_{B},B_{\mathrm{ph}}^{i}]=0$ and hence $\sum_{i}B_{\mathrm{ph}}^{i}|\phi
_{s}\rangle =s|\phi _{s}\rangle $. Under these assumptions we have that $%
[H_{0},A_{+}B_{\mathrm{ph}}^{i}]=0$, and hence the effect of the pulse
operator (\ref{eq:U}) on the initially uncorrelated system-bath state can be
written as
\begin{eqnarray}
\rho _{\mathrm{SB}}(\tau _{L},0) &\approx &U(\tau _{L},0)(|\psi \rangle
\langle \psi |\otimes \rho _{B})U^{\dag }(\tau _{L},0)  \nonumber
\label{state1} \\
&=&\sum_{s}p_{s}e^{-i\tau _{L}\overline{H_{0}}}e^{-i\tau _{L}\overline{%
\Gamma _{+}}sA_{+}}  \nonumber \\
&&\times (|\psi \rangle \langle \psi |\otimes |\phi _{s}\rangle \langle \phi
_{s}|)e^{i\tau _{L}\overline{H_{0}}}e^{i\tau _{L}\overline{\Gamma _{+}}%
sA_{+}}\;.  \nonumber \\
\end{eqnarray}%
Using the expression (\ref{state}) for $|\psi \rangle $, we find, after
tracing out the bath, that the state of the field modes at the final time $%
\tau _{L}$ becomes
\begin{eqnarray}  \label{r-out}
\rho (\tau _{L}) &=&\frac{1}{P^{2}}\sum_{s}p_{s}\sum_{n,m}\zeta ^{n+p}\zeta
^{\ast (m+p)}\sqrt{\frac{(n+p)!(m+p)!}{n!m!}}  \nonumber \\
&&\times e^{-i\tau _{L}[\omega _{\mathrm{tot}}+2\Gamma
_{+}s](n-m)}|n,n+p\rangle \langle m,m+p|\;,  \nonumber \\
\end{eqnarray}%
where $\omega _{\mathrm{tot}}=\omega _{a}+\omega _{b}$. The fidelity of this
state can be calculated as
\begin{eqnarray}
&&F(\tau _{L}) =\langle \psi |\rho (\tau _{L})|\psi \rangle  \nonumber
\label{fid} \\
&=&\frac{1}{P^{4}}\sum_{s}p_{s}\left\vert \sum_{n}|\zeta |^{2(n+p)}\frac{%
(n+p)!}{n!}e^{-i\tau _{L}n(\omega _{\mathrm{tot}}+2\Gamma _{+}s)}\right\vert
^{2}\;,  \nonumber \\
\end{eqnarray}%
which becomes unity if
\begin{equation}
\tau _{L}=l\pi /\{\omega _{\mathrm{tot}}+2\Gamma _{+}s\}\;,  \label{cond}
\end{equation}%
where $l$ is an integer. In this expression $\tau _{L}$ is, of course a
function of the summation variable $s$, while in reality there is only a
single $\tau _{L}$. However, we observe that there are two physical limits
where this dependence of $\tau _{L}$ on $s$ disappears. Namely, if $\omega _{%
\mathrm{tot}}\ll \Gamma _{+}$ then for $l=s$ we find that $\tau $ becomes $s$%
-independent. In the opposite limit of very weak collective dephasing we
also find that $\tau $ is $s$-independent (there is a physical upper limit
on $s$ in the Gibbs state $\rho _{B}=\sum_{s}p_{s}|\phi _{s}\rangle \langle
\phi _{s}|$). Both limits can be realized by controlling the field modes $%
\omega _{a,b}$. Alternatively, in the low temperature limit the sum over $s$
in Eq.~(\ref{fid}) is dominated by the vibrational ground state, and again
the fidelity is unity if $\tau =l\pi /\omega _{\mathrm{tot}}$.

While the \emph{fidelity} is generally reduced and recovers its initial
value of unity only under certain conditions, we next show that the
non-Gaussian entanglement can be preserved provided one knows the value of $%
\Gamma _{+}$. At the exit end of the fiber, the state of the two modes is
given by $\rho (\tau _{L},0)$ [Eq.~(\ref{r-out})]. Taking the partial
transpose over the mode $b$ in this state, we obtain:
\begin{eqnarray}
\rho (\tau _{L})^{\mathrm{T_{b}}} &=&\frac{1}{P^{2}}\sum_{s}p_{s}\sum_{n,m}%
\zeta ^{n+p}\zeta ^{\ast (m+p)}\sqrt{\frac{(n+p)!(m+p)!}{n!m!}}  \nonumber
\label{rhoa} \\
&&\times e^{-i\tau _{L}[\omega _{\mathrm{tot}}+2\Gamma
_{+}s](n-m)}|n,m+p\rangle \langle m,n+p|\;,  \nonumber \\
\end{eqnarray}%
The eigenvalues of this matrix are given by
\begin{eqnarray}
\lambda _{nn} &=&\frac{1}{P^{2}}|\zeta |^{2(n+p)}\frac{(n+p)!}{n!}%
\;\;\;\forall n\;,  \nonumber \\
\lambda _{nm} &=&\pm \frac{1}{P^{2}}v_{nm}(\tau _{L}\Gamma _{+}) |\zeta
|^{n+m+2p}  \nonumber \\
&\times &\sqrt{\frac{(n+p)!(m+p)!}{n!m!}}\;\;\;\forall n,m\neq n,
\end{eqnarray}%
where we have defined the \emph{phononic visibility factor} as
\begin{equation}
v_{nm}(x)\equiv \left\vert \sum_{s}p_{s}e^{-2ixs(n-m)}\right\vert .
\end{equation}%
For a harmonic oscillator bath this factor can be evaluated analytically,
using $E_{s}=\hbar \Omega (s+1/2)$:%
\begin{eqnarray}
v_{nm}(x)&=&\frac{1}{Z}e^{-\hbar \Omega /2k_{B}T}\left\vert
\sum_{s=0}^{\infty }e^{-[\hbar \Omega /k_{B}T+2ix(n-m)]s}\right\vert
\nonumber \\
&=&\frac{1}{2Z}\left\vert \frac{1}{\sinh (\hbar \Omega /2k_{B}T+ix(n-m))}%
\right\vert ,
\end{eqnarray}%
which is an oscillatory function of its argument $x$, with period $\pi
/(n-m) $.

The negativity at the fiber's end becomes:
\begin{equation}
\mathcal{N}=\frac{1}{P^{2}}\sum_{n=0}^{\infty }\sum_{m\neq n}^{\infty
}v_{nm}(\tau _{L}\Gamma _{+})|\zeta |^{n+m+2p}\sqrt{\frac{(n+p)!(m+p)!}{n!m!}%
}\;,  \label{neg2}
\end{equation}%
Comparing this expression to Eq.~(\ref{neg}), it is apparent that the
negativity is reduced due to the phononic visibility factor. The condition
for this factor to become unity is:
\begin{equation}
\tau =\pi /\Gamma _{+},
\end{equation}%
which can be satisfied provided one knows the value of $\Gamma _{+}$. One
way to extract this value is, in fact, to apply our bang-bang protocol and
to test the fidelity of the state via quantum state tomography (see \cite%
{ByrdLidar:02} for a more general method relating BB\ to tomography). Also
note that in the low temperature limit, where the sum over $s$ involves only
a small number of terms, the phononic visibility factor will still be close
to unity. On the other hand, it is clear that if neither condition is met ($%
\tau \neq \pi /\Gamma _{+}$ and high temperature) then the rapidly
oscillating terms in the phononic visibility factor will destructively
interfere and cause entanglement loss.

Note also that in absence of any BB control, the initial state $|\psi\rangle$
evolves through the interaction Hamiltonian (\ref{ham}). This leads to
different combinations of the photon numbers in the two modes due to energy
exchange with the molecular bath. Thus the initial entanglement is destroyed
at a length scale $\Delta_c$, corresponding to the dissipation time-scale $%
\tau_c$ of the molecular bath.

\section{Effect of fiber inhomogeneity on decoherence}

\label{inhom}

So far we have assumed that the average Hamiltonians over two successive
segments of length $\Delta $ are equal [Eq.~(\ref{eq:approxH})]. However,
due to inhomogeneity inside the fiber, the average Hamiltonian differs
between segments. In this section, we show that this fluctuation of the
average Hamiltonian leads to dissipation of the field and thus sets an upper
limit to the value of $\Delta $. We follow and improve the method described
in Appendix A of \cite{lidar}.

\subsection{A Gaussian fluctuations model}

The inhomogeneity in the fiber may arise due to nonuniform number density of
the molecules or slow time-dependence of the fiber properties. In view of
this, we modify the assumption of homogeneity to read
\begin{equation}
H((k-1)\Delta )=H(k\Delta )+\delta (H_{B})_{k}+\delta (H_{I})_{k},
\label{fluc}
\end{equation}%
where the operator-valued fluctuations $\delta (H_{B})_{k},\delta
(H_{I})_{k} $ in the Hamiltonian are independent. Here $\delta (H_{B})_{k}$
\ is the fluctuation in the bath-only Hamiltonian, and $\delta (H_{I})_{k}$
\begin{eqnarray}
\delta (H_{I})_{k} &=&\sum_{i}\delta g_{i}^{k}(a^{\dag }bB_{i}+ab^{\dag
}B_{i}^{\dag })_{k}  \nonumber \\
&+&\epsilon \sum_{i}(\delta \Gamma _{a}^{i}n_{a}+\delta \Gamma
_{b}^{i}n_{b})_{k}(B_{\mathrm{ph}}^{i})_{k}  \label{fluc_hamilt}
\end{eqnarray}%
is the operator-valued correction to the interaction Hamiltonian in the $k$%
th segment, where $\delta g_{i}^{k}$ and $\delta (\Gamma _{l}^{i})_{k}$ ($%
l=a,b$) are the fluctuations in the coupling coefficients in the $k$th
segment, and $\epsilon \ll 1$ is a proportionality constant defining the
strength of the fluctuations. We assume that $\epsilon \sim \tau $. These
fluctuations lead to losses inside the fiber. However, in an amorphous
silica fiber, the loss due to Rayleigh scattering (i.e., due to terms
containing $n_{a}$ and $n_{b}$) is much greater than the loss due to other
mechanisms (e.g., due to Raman scattering, i.e., due to terms containing $%
a^{\dag }b$ and $ab^{\dag }$) \cite{gpa}. With this in mind, we neglect the fluctuations in the
Raman terms and henceforth consider only the effect of the fluctuations in $(\Gamma
_{a,b}^{i})_{k}$. Thus, in the interaction picture with respect to the energy term $H_0+\delta
H_B(t)$, the interaction Hamiltonian corresponding to the fluctuations in the $k$th segment of the
fiber can be written as
\begin{equation}
\delta H_{I}(t)=\epsilon \sum_{i}(f_{a}^{i}(t)n_{a}+f_{b}^{i}(t)n_{b})\;,
\end{equation}%
where $f_{l}^{i}(t)\equiv \delta \Gamma _{l}^{i}(t)B_{\mathrm{ph}}^{i}$ ($%
l=a,b$) are the operator-valued fluctuations for the $i$th molecule, and we
have replaced the $k$-dependence with time-dependence, in the joint limit $%
\tau \rightarrow 0$ and total number $N$ of segments large, such that $\tau
_{L}=N\tau $ is finite. Under the action of this Hamiltonian $\delta
H_{I}(t) $, the evolution of the field state can be described, in the Born
approximation, by the following equation \cite{scully}:
\begin{eqnarray}
\dot{\rho}_{F} &=&-i\mathrm{Tr}_{B}[\delta H_{I}(t),\rho _{F}(0)\otimes \rho
_{B}(0)]  \nonumber \\
&-&\mathrm{Tr}_{B}\int_{0}^{t}[\delta H_{I}(t),[\delta H_{I}(t^{\prime
}),\rho _{F}(t^{\prime })\otimes \rho _{B}(0)]],  \label{fluc_eqn}
\end{eqnarray}%
where the terms up to the second order of $\epsilon (\sim \tau )$ have been
considered (as $\epsilon \ll 1$). Now note that since $\Delta \ll L$, the
number of segments of length $\Delta $ is much larger than unity. In this
limit, the fluctuations $f_{l}^{i}(t)$ can be considered as described by
Gaussian operators \cite{lidar}. Then it is reasonable to assume that the
two-time correlations of the form $\Gamma (t,t^{\prime })\equiv \left\langle
\left( \sum_{i}f_{l}^{i}(t)\sum_{j}f_{l^{\prime }}^{j}(t^{\prime })\right)
\right\rangle _{B}$ ($l,l^{\prime }=a,b$) between these fluctuations do not
depend on $a,b$ due to the symmetry properties of Gaussian operators, while
the mean $\langle \sum_{i}f_{l}^{i}(t)\rangle _{B}$ vanishes. Here $\langle
X\rangle _{B}=\mathrm{Tr}_{B}(\rho _{B}(0)X)$ denotes the average of any
operator $X$ over the bath. Taking this average over Eq.~(\ref{fluc_eqn})
leads to the vanishing of the term linear in $\delta H_{I}(t)$, while using
the expression for $|\psi \rangle $ [Eq.~(\ref{state})] we obtain from the
integral term:%
\begin{eqnarray}
\dot{\rho}_{F} &=&-\frac{4\epsilon ^{2}}{P^{2}}\int_{0}^{t}dt^{\prime
}\Gamma (t,t^{\prime })\sum_{n,m}\zeta ^{n+p}\zeta ^{\ast (m+p)}  \nonumber
\\
&&\sqrt{\frac{(n+p)!(m+p)!}{n!m!}}(n-m)^{2}|n,n+p\rangle \langle m,m+p|.
\nonumber \\
&&
\end{eqnarray}%
This can be solved for the matrix elements of $\rho _{F},$ yielding
\begin{eqnarray}
(\rho _{F})_{n,n+p;m,m+p}(\tau _{L}) &=&\exp [-4\epsilon
^{2}(n-m)^{2}\int_{0}^{\tau _{L}}dt\int_{0}^{t}dt^{\prime }  \nonumber \\
&&\Gamma (t,t^{\prime })](\rho _{F})_{n,n+p;m,m+p}(0),  \nonumber \\
&&  \label{decay_sol}
\end{eqnarray}%
where the initial state is
\begin{equation}
(\rho _{F})_{n,n+p;m,m+p}(0)=\frac{1}{P^{2}}\zeta ^{n+p}\zeta ^{\ast (m+p)}%
\sqrt{\frac{(n+p)!(m+p)!}{n!m!}}.
\end{equation}

Let us assume that the molecular bath is in thermal equilibrium at
temperature $T$. If $T\rightarrow 0$, most of the molecules reside in the
ground electronic states. However, molecules can be distributed in all the
vibronic modes corresponding to the ground states. The degeneracy of these
vibronic states can be lifted by phonon absorption due to molecular
collisions. In view of this, we treat the molecules as bosons inside a
fiber. As implied by our discussion in the previous section, the dephasing
operator $\sum_{i}B_{\mathrm{ph}}^{i}$ can be written as $(B^{\dag }B)_{%
\mathrm{vib}}$, where $(B)_{\mathrm{vib}}$ is the annihilation operator
corresponding to these vibronic states. Thus, the interaction of the field
modes with the molecules can be described by the so-called independent
oscillator (IO) model \cite{ford} of boson-boson interactions. In this
model, each of the oscillators of the \emph{passive} molecular bath is \emph{%
linearly} coupled to the system oscillator. This interaction is governed by
two terms: (i) The Gaussian fluctuations $\sum_{i}f_{l}^{i}(t)$ of the bath
operators, and (ii) a memory function $\mu (t)$ of the bath operators, that
vanishes at negative times. If the correlation between fluctuations at
different times vanishes, the interaction reduces to a Markovian process.
However, in general, this correlation is a non-trivial function of time and
thus corresponds to a non-Markovian process. It can be shown that the
symmetric autocorrelation of $\sum_{i}f_{l}^{i}(t)$ satisfies \cite{ford}:
\begin{eqnarray}
&&\frac{1}{2}\left\langle \sum_{i}f_{l}^{i}(t)\sum_{j}f_{l^{\prime
}}^{j}(t^{\prime })+\sum_{j}f_{l^{\prime }}^{j}(t^{\prime
})\sum_{i}f_{l}^{i}(t)\right\rangle  \nonumber  \label{corr} \\
&=&\frac{1}{\pi }\int_{0}^{\infty }\omega \coth \left( \frac{\hbar \omega }{%
2k_{B}T}\right) \cos [\omega (t-t^{\prime })]  \nonumber \\
&&\;\;\;\;\;\mathrm{Re}[\tilde{\mu}(\omega +i0^{+})]d\omega \;,
\end{eqnarray}%
while correlations of an odd number of factors of $\sum_{i}f_{l}^{i}(t)$
vanish. Here $\tilde{\mu}(z)$ is the Fourier transform of the memory
function $\mu (t)$. Due to the Gaussian property of $\sum_{i}f_{l}^{i}(t)$,
as discussed before, the above equals $\langle
\sum_{i}f_{l}^{i}(t)\sum_{j}f_{l^{\prime }}^{j}(t^{\prime })\rangle $. The
memory function is independent of the potential and the properties of the
system and only depends upon the coupling strengths of the field operators
with the bath. In the IO model, the coupling strength is an even function of
the oscillator frequency and is of the form $m_{j}\omega _{j}^{2}$, where $%
m_{j}$ is the mass of the $j$th oscillator and the $\omega _{j}$ is the
frequency of the $j$th oscillator mode. The spectral distribution of the
memory function is in the Ohmic class and can be written as
\begin{equation}
\mathrm{Re}[\tilde{\mu}(\omega +i0^{+})]\approx \pi \omega ^{2}e^{-\omega
/\omega _{c}}\sum_{j}m_{j}\;,  \label{muomega}
\end{equation}%
Here $\omega _{c}$ is the cut-off frequency of the molecular bath modes, and
is introduced such that for large frequencies, the memory function does not
blow up. This is in conformity with the passivity condition of the molecular
bath, which also requires that the bath modes must have an infinite spectrum
and the memory function must not be a singular function of $\omega $ \cite%
{ford}. Thus using Eqs.~(\ref{corr}) and (\ref{muomega}), we can write the
following expression for the rate of dissipation at time $\tau _{L}$:
\begin{eqnarray}
\Gamma (\tau _{L}) &=&\int_{0}^{\tau _{L}}dt\int_{0}^{t}dt^{\prime }\Gamma
(t,t^{\prime })  \nonumber  \label{gamma2} \\
&=&\int_{0}^{\infty }d\omega \omega ^{3}e^{-\omega /\omega _{c}}\coth \left[
\frac{\hbar \omega }{2k_{B}T}\right] \left\{ \frac{2}{\omega ^{2}}\sin
^{2}\left( \frac{\omega \tau_L}{2}\right) \right\} \;,  \nonumber \\
&&
\end{eqnarray}%
where we have normalized the memory function (\ref{muomega}) in units of
mass. For low temperature ($T\rightarrow 0$), when the quantum fluctuations
dominate the thermal fluctuations, one can find the actual loss figure as
\begin{equation}
\Gamma (\tau _{L})\approx \omega _{c}^{2}\frac{x^{2}(3+x^{2})}{(1+x^{2})^{2}}%
,\;x=\omega _{c}\tau _{L}\;.  \label{loss}
\end{equation}%
Using the results for the unitary evolution (\ref{r-out}) and the decay of
the density matrix elements given by (\ref{decay_sol}), we find the
following expression for the density matrix of the field modes at time $\tau
_{L}$:
\begin{eqnarray}
\rho (\tau _{L})_{\mathrm{diss}} &=&\frac{1}{P^{2}}\sum_{s}p_{s}\sum_{n,m}%
\zeta ^{n+p}\zeta ^{\ast (m+p)}\sqrt{\frac{(n+p)!(m+p)!}{n!m!}}  \nonumber \\
&\times &e^{-i\tau _{L}(\omega _{\mathrm{tot}}+2\Gamma
_{+}s)(n-m)}e^{-4\epsilon ^{2}\Gamma (\tau _{L})(n-m)^{2}}  \nonumber \\
&&\;\;\;|n,n+p\rangle \langle m,m+p|\;.
\end{eqnarray}%
The eigenvalues of the matrix obtained by taking transpose of the $b$ mode
in the above density matrix are given by
\begin{eqnarray}
\lambda _{nn} &=&\frac{1}{P^{2}}|\zeta |^{2(n+p)}\frac{(n+p)!}{n!}%
\;\;\;\forall n\;,  \nonumber \\
\lambda _{nm} &=&\pm \frac{1}{P^{2}}\left\vert \sum_{s}p_{s}e^{-2i\tau
_{L}\Gamma _{+}s(n-m)}\right\vert |\zeta |^{n+m+2p}  \nonumber \\
&&\times \sqrt{\frac{(n+p)!(m+p)!}{n!m!}}e^{-4\epsilon ^{2}\Gamma (\tau
_{L})(n-m)^{2}}\;\;\;\forall n,m\neq n\;.  \nonumber \\
&&
\end{eqnarray}%
Thus in the low temperature limit, we find the following expression
for the negativity:
\begin{eqnarray}
\mathcal{N}_{\mathrm{diss}} &=&\frac{1}{P^{2}}\sum_{n,m\neq n}|\zeta
|^{n+m+2p}\sqrt{\frac{(n+p)!(m+p)!}{n!m!}}  \nonumber \\
&&\times e^{-4\epsilon ^{2}\Gamma (\tau _{L})(n-m)^{2}},  \label{neg_diss}
\end{eqnarray}%
where $\Gamma (\tau _{L})$ is given by Eq.~(\ref{loss}). Clearly the
negativity decreases as the field propagates through the fiber. Using the
expression (\ref{loss}) for $\Gamma (\tau _{L})$, we find that for $x\gg 1$
(for $\tau _{L}\gg 1/\omega _{c}$, i.e., on a time-scale much larger than
the bath correlation time), the negativity does not decrease further and
saturates to (\ref{neg_diss}) with $\Gamma (\tau _{L})$ replaced by $\omega
_{c}^{2}$. This saturation is due to the fact that $1/\omega_c$ is the
timescale over which information about the system state spreads in the bath;
for times much longer than this the bath is effectively stationary and no
more damage to entanglement in the system is possible via the bath. We next
extract a distance scale for the phase shifter separations from these
considerations.

\subsection{Numerical estimates of the inter phase-shifter distance}

The result (\ref{loss}) due to fluctuation given by (\ref{fluc_hamilt})
leads to an estimate of the spatial separation $\Delta $ between two
successive phase-shifters. Note that for $n=m$ (i.e., for diagonal terms of
the density matrix) this fluctuation does not lead to any dissipation (or
loss of entanglement), as is clear from (\ref{decay_sol}). But the larger
is $|n-m|$, the higher is this dissipation. In other words, the largest
off-diagonal contribution to the negativity comes from the terms with $%
|n-m|=1$, which therefore are sufficient to give us the desired estimate of $%
\Delta $. If we allow an error probability $\delta (\tau _{L})$ over the
length of the fiber, then a sufficient condition for the present result to
be useful is
\begin{equation}
e^{-4\epsilon ^{2}\Gamma (\tau _{L})}>1-\delta (\tau _{L})\;.
\end{equation}%
Rewriting the above using (\ref{loss}), we have
\begin{eqnarray}
\Delta  &<&\left( \frac{v^{2}(1+x^{2})^{2}}{4\omega _{c}^{2}x^{2}(3+x^{2})}%
\ln [1-\delta (\tau _{L})]^{-1}\right) ^{1/2}  \nonumber \\
&&\overset{x\rightarrow \infty }{\longrightarrow }\frac{v}{2\omega _{c}}\ln
[1-\delta (\tau _{L})]^{-1/2},  \label{limit}
\end{eqnarray}%
where we have used $\epsilon =\tau $. The limiting value is attainable by
fixing $\Delta $ and letting the number of phase shifters $N$ become very
large (long fiber), with $\omega _{c}$ given and fixed. \emph{This sets an
upper bound on the applicable value of }$\Delta $\emph{, which is
essentially the ratio of the speed of light in the fiber to the bath
high-frequency cutoff.}

In the following, we present a numerical estimate in a realistic situation,
e.g., for an optical fiber with an amorphous silica core. The inhomogeneity
in silica leads to fluctuation as described above and thus decoheres the
input fields.

The Debye temperature $\Theta=\hbar\omega_c/k_B$ of crystalline silica is
342 K, where $\omega_c$ is the maximum phonon frequency (frequency
``cut-off'') allowed inside the crystal. Thus, the lifetime $\tau_c$ of
phonons becomes of the order of $1/\omega_c$. On the other hand, in
amorphous solids the Debye temperature and the lifetime of phonons are not
well defined. However, at low temperatures $T\ll \Theta$, there exist
certain empirical relations between them \cite{anderson,reynolds}. For
example, at $T=0.2$ K, which corresponds to a phonon frequency of $%
2.62\times 10^{10}$ Hz, the life-time of phonons is of the order of $%
10^{-10} $ s. We consider this frequency of phonons as maximum frequency $%
\omega_c$ allowed inside the fiber at $T=0.2$K.

It has been shown that long distance distribution of entangled states of
two qubits over a noisy quantum channel can be achieved using entanglement
purification protocols \cite{bennett,deutsch}. These protocols can be
improved in terms of the requirement of physical resources as well as the
error threshold, if one uses quantum repeaters \cite{dur}. It has been shown
that for an error probability $\sim 0.01$ (which is much larger than the
error threshold for fault-tolerant computation using single qubits) inside
the communication channel, quantum purification protocols work well, when
combined with quantum repeaters. Although, as explained in the introduction,
the setup of our problem is quite different from that of quantum repeaters,
for the sake of concreteness we use the threshold figure from that scenario
and conservatively consider the case when the maximum error probability $%
\delta (\tau _{L})$ allowed through the fiber is $5\times 10^{-2}$.

We consider a multimode fiber of length $L=1$ km. The time of propagation of
the fields through the fiber is $\tau _{L}=5.33\times 10^{-6}$ s, where $%
v=c/n_{g}$, $n_{g}=1.6$ being the effective group index of the field through
the fiber. Using the parameters discussed above, we find from Eq.~(\ref%
{limit}) $\Delta \lesssim 0.8$ mm. This means that the field sees the phase
shifters at a time interval $\tau $ of $4.325\times 10^{-12}$sec which, as
required for the BB protocol, is \emph{much smaller} than the time-scale for
bath dissipation, $\tau _{c}=10^{-10}$sec. We show in Fig. 2 how the
negativity (\ref{neg_diss}) varies inside the fiber for the parameters
discussed here. We find that the negativity becomes constant after a certain
length scale inside the fiber, as discussed before. This is because, in the
presence of bang-bang control, the effective contribution of bath
fluctuations to the negativity vanishes at a time-scale when the bath
correlation vanishes.

\begin{figure}[tbp]
\vspace{0.8cm} \centerline{\scalebox{0.3}{\includegraphics{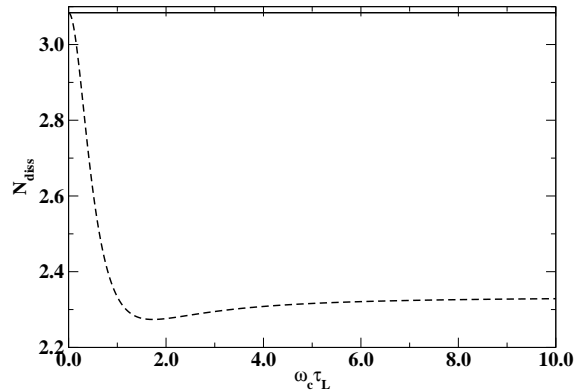}}} \caption{Variation of the
negativity ${\cal N}_{\mathrm{diss}}$ with time in units of the bath high-frequency cut-off,
$\protect\omega_c\tau_L$ for a fixed squeezing parameter $|\protect\zeta|=0.5$ and $p=1$ accounting
for (dashed line) and not accounting for (solid line) the bath fluctuations. The
other parameters are $\protect\omega_c=2.62\times 10^{10}$ Hz, $\protect\tau%
=4.325\times 10^{-12}$ s, and error probability $\protect\delta=5\times
10^{-2}$.}
\label{fig2}
\end{figure}

\section{Conclusions}

In conclusion, we have discussed in detail, how one can preserve
entanglement in a class of continuous variable non-Gaussian states against
decoherence caused by coupling to a bosonic bath. Specifically, we have
considered transmission of an entangled state of two bosonic modes through
an optical fiber and developed a hybrid approach combing decoherence-free
subspaces and bang-bang control to sustain the entanglement. We described
the non-Markovian interaction with the bosonic bath consisting of molecules
in the fiber and provided a detailed estimate of the relevant parameters to
implement our approach in a realistic fiber. It turns out that to achieve a
loss figure of $<5\times 10^{-2}$ in a 1 km fiber, phase shifters should be
placed about 1 mm apart. This appears to be a technologically feasible
requirement. Hence we expect that the method proposed here will become
a useful tool in the effort to transmit non-Gaussian entangled states
over optical fibers.


\begin{thebibliography}{99}
\bibitem{geo} V. Giovannetti, S. Guha, S. Lloyd, L. Maccone,
  J. H. Shapiro, and H. P. Yuen, Phys. Rev. Lett. \textbf{92}, 027902 (2004).

\bibitem{bartlett} S. D. Bartlett, B. C. Sanders, S. L. Braunstein,
  and K. Nemoto, Phys. Rev. Lett. \textbf{88}, 097904 (2002);
  S. D. Bartlett and B. C. Sanders, Phys. Rev. Lett. \textbf{89},
  207903 (2002).

\bibitem{eisert} J. Eisert, S. Scheel, and M. B. Plenio, Phys. Rev. Lett.
\textbf{89}, 137903 (2002).

\bibitem{biswas} G. S. Agarwal and A. Biswas, New J. Phys. \textbf{7}, 211
(2005).

\bibitem{vogel} E. Shchukin and W. Vogel, Phys. Rev. Lett. \textbf{95},
230502 (2005).

\bibitem{hillery} M. Hillery and M. S. Zubairy, Phys. Rev. Lett. \textbf{96},
050503 (2006).

\bibitem{bennett} C. H. Bennett, G. Brassard, S. Popescu, B. Schumacher, J.
A. Smolin, and W. K. Wootters, Phys. Rev. Lett. \textbf{76}, 722 (1996); C. H. Bennett, D. P.
DiVincenzo, J. A. Smolin, and W. K. Wootters, Phys. Rev. A \textbf{54}, 3824 (1996).

\bibitem{deutsch} D. Deutsch, A. Ekert, R. Jozsa, C. Macchiavello, S. Popescu, and A.
Sanpera, Phys. Rev. Lett. \textbf{77}, 2818 (1996).

\bibitem{dur} H.-J. Briegel, W. D\"{u}r, J.I. Cirac, P. Zoller, Phys. Rev.
Lett. \textbf{81}, 5932 (1998); W. D\"ur, H.-J. Briegel, J. I. Cirac, and P. Zoller, Phys. Rev. A
\textbf{59}, 169 (1999).

\bibitem{WuLidarSchneider:04} L.-A. Wu, D. A. Lidar, and S. Schneider, Phys.
Rev. A \textbf{70}, 032322 (2004).

\bibitem{Compagno:04} G. Compagno, A. Messina, H. Nakazato, A. Napoli, M.
Unoki, and K. Yuasa, Phys. Rev. A \textbf{70}, 052316 (2004).

\bibitem{dfs} {L.-M Duan and G.-C. Guo}, Phys. Rev. Lett. {\bf 79},
  1953 (1997); P. Zanardi and M. Rasetti,
  Phys. Rev. Lett. \textbf{79}, 3306
(1997); D. A. Lidar, I. L. Chuang, and K. B. Whaley, Phys. Rev. Lett.
\textbf{81}, 2594 (1998); D. A. Lidar and K. B. Whaley, \textit{Irreversible
Quantum Dynamics\/}, F. Benatti and R. Floreanini (Eds.), p. 83 (Springer Lecture Notes in Physics,
\textbf{622}, Berlin, 2003).

\bibitem{BB} L. Viola and S. Lloyd, Phys. Rev. A \textbf{58}, 2733
  (1998); P. Zanardi, Phys. Lett. A {\bf 258}, 77 (1999); L.-M Duan
  and G. C. Guo, Phys. Lett. A {\bf 261}, 139 (1999); L. Viola,
  E. Knill, and S. Lloyd, Phys. Rev. Lett. \textbf{82}, 2417 (1999);
  M. S. Byrd and D. A. Lidar, Quant. Info. Proc. \textbf{1}, 19
  (2002); P. Facchi, D.A. Lidar, and S. Pascazio, Phys. Rev. A {\bf 69},
    032314 (2004).

\bibitem{lidar} L.-A. Wu and D. A. Lidar, Phys. Rev. A \textbf{70}, 062310
(2004).

\bibitem{byrd_lidar} M. S. Byrd and D. A. Lidar, Phys. Rev. Lett. \textbf{89}%
, 047901 (2002).

\bibitem{LidarWu-ERD} D. A. Lidar and L.-A. Wu, Phys. Rev. A \textbf{67},
032313 (2003).

\bibitem{tara_thermal} G. S. Agarwal and K. Tara, Phys. Rev. A \textbf{43},
492 (1991).

\bibitem{gsapuri} G. S. Agarwal, R. R. Puri, and R. P. Singh, Phys. Rev. A
\textbf{56}, 4207 (1997).

\bibitem{gsa_qo} G. S. Agarwal, Quant. Opt. \textbf{2}, 1 (1990); G. M.
D'Ariano, P. Kumar, C. Macchiavello, L. Maccone, and N. Sterpi, Phys. Rev.
Lett. \textbf{83}, 2490 (1999).

\bibitem{yamamoto} K. Watanabe and Y. Yamamoto, Phys. Rev. A \textbf{38},
3556 (1998).

\bibitem{bellini} A. Zavatta, S. Viciani, and M. Bellini, Science \textbf{306%
}, 660 (2004).

\bibitem{grangier} R. Garc\'ia-Patr\'on, J. Fiur\'a\u sek, N. J. Cerf, J.
Wenger, R. Tualle-Brouri, and P. Grangier, Phys. Rev. Lett. \textbf{93},
130409 (2004); R. Garc\'ia-Patr\'on, J. Fiur\'a\u sek, and N. J. Cerf, Phys.
Rev. A \textbf{71}, 022105 (2005).

\bibitem{peres} A. Peres, Phys. Rev. Lett. \textbf{77}, 1413 (1996); M.
Horodecki, P. Horodecki, and R. Horodecki, Phys. Lett. A \textbf{223}, 1
(1996).

\bibitem{vidal} G. Vidal and R. F. Werner, Phys. Rev. A \textbf{65}, 032314
(2002).

\bibitem{cohen} C. Cohen-Tannoudji,, J. Dupont-Roc, and G. Grynberg, \textit{%
Atom-Photon Intercations: Basic Processes and Applications} (Wiley, New
York, 1998), p. 101.

\bibitem{Reinsch:00} {M.W. Reinsch}, J. Math. Phys. \textbf{41}, 2434 (2000).

\bibitem{ByrdLidar:02}  M.S. Byrd and D.A. Lidar, Phys. Rev. A \textbf{67},
012324 (2003).

\bibitem{gpa} G. P. Agarwal, \emph{Fiber-optic Communication Systems}
(Wiley, New York, 1992), Sec. 2.5 and Fig. 2.15.

\bibitem{scully} M. O. Scully and M. S. Zubairy, \emph{Quantum Optics}
(Cambridge, Cambridge, 1997), Sec. 8.1.

\bibitem{ford} G. W. Ford, J. T. Lewis, and R. F. O'Connell, Phys. Rev.
Lett. \textbf{55}, 2273 (1985); Phys. Rev. A \textbf{37}, 4419 (1988).

\bibitem{anderson} J. J. Freeman and A. C. Anderson, Phys. Rev. B \textbf{34}%
, 5684 (1986).

\bibitem{reynolds} C. L. Reynolds, J. Non-Cryst. Solids \textbf{37}, 125
(1980).

\end{thebibliography}
\end{document}